\documentclass[a4paper,11pt,centertags,psamsfonts]{amsart}
\usepackage{times,amssymb}
\usepackage[mathcal]{euscript}

\newfont{\timitfont}{pplbi7t scaled 1000}
\newcommand{\vit}{\mbox{\timitfont{v}}}
\newfont{\timitfontsmall}{pplbi7t scaled 800}
\newcommand{\vitsmall}{\mbox{\timitfontsmall{v}}}

\DeclareSymbolFont{operators}   {OT1}{ptmcm}{m}{n}
\DeclareSymbolFont{letters}     {OML}{ptmcm}{m}{it}
\DeclareSymbolFont{symbols}     {OMS}{pzccm}{m}{n}
\DeclareSymbolFont{largesymbols}{OMX}{psycm}{m}{n}

\DeclareMathAlphabet{\mathrm}{OT1}{ptm}{m}{n}

\DeclareSymbolFont{ER}{U}{eur}{m}{n}
\DeclareSymbolFont{SY}{U}{psy}{m}{n}

\DeclareMathSymbol{,}{\mathpunct}{SY}{'054}
\DeclareMathSymbol{.}{\mathpunct}{SY}{'056}
\DeclareMathSymbol{:}{\mathpunct}{SY}{'072}

\DeclareMathSymbol{(}{\mathopen}{SY}{'050}
\DeclareMathSymbol{)}{\mathclose}{SY}{'051}

\DeclareMathSymbol{+}{\mathbin}{SY}{'053}
\DeclareMathSymbol{-}{\mathbin}{SY}{'055}
\DeclareMathSymbol{=}{\mathbin}{SY}{'075}
\DeclareMathSymbol{<}{\mathbin}{SY}{'074}
\DeclareMathSymbol{>}{\mathbin}{SY}{'076}
\DeclareMathSymbol{\leq}{\mathbin}{SY}{'243}
\DeclareMathSymbol{\geq}{\mathbin}{SY}{'263}
\DeclareMathSymbol{\nneq}{\mathbin}{SY}{'271}
\DeclareMathSymbol{\in}{\mathbin}{SY}{'316}
\DeclareMathSymbol{\nnotin}{\mathbin}{SY}{'317}
\DeclareMathSymbol{\times}{\mathbin}{SY}{'264}
\DeclareMathSymbol{\pm}{\mathbin}{SY}{'261}
\DeclareMathSymbol{\subset}{\mathbin}{SY}{'314}
\DeclareMathSymbol{\supset}{\mathbin}{SY}{'311}
\DeclareMathSymbol{\subseteq}{\mathbin}{SY}{'315}
\DeclareMathSymbol{\supseteq}{\mathbin}{SY}{'312}

\DeclareMathSymbol{/}{\mathord}{SY}{'057}
\DeclareMathSymbol{|}{\mathord}{SY}{'174}
\DeclareMathSymbol{\ast}{\mathord}{SY}{'052}
\DeclareMathSymbol{\perp}{\mathord}{SY}{'136}

\renewcommand{\neq}{\nneq}
\renewcommand{\notin}{\nnotin}

\newcommand{\Z}{\mathbb{Z}}
\newcommand{\R}{\mathbb{R}}

\newcommand{\N}{\mathbb{N}}
\newcommand{\supp}{{\ensuremath{\rm supp}}}

\DeclareMathOperator*{\slim}{s-lim}
\newcommand{\Ran}{\mathrm{Ran}}

\newcommand{\cD}{\mathcal{D}}
\newcommand{\cH}{\mathcal{H}}
\newcommand{\cQ}{\mathcal{Q}}
\newcommand{\cR}{\mathcal{R}}
\newcommand{\cS}{\mathcal{S}}

\newtheorem{theorem}{Theorem}[section]{\bf}{\it}

\newcommand{\x}{\mathbf{x}}
\newcommand{\p}{\mathbf{p}}
\newcommand{\e}{\mathbf{e}}
\renewcommand{\v}{\mathbf{v}}

\newtheorem{proposition}[theorem]{Proposition}{\bf}{\it}
\newtheorem{corollary}[theorem]{Corollary}{\bf}{\it}
\newtheorem{lemma}[theorem]{Lemma}{\bf}{\it}

\title[Scattering By Rotating Potentials]{Energy Transfer in Scattering
by Rotating Potentials}

\author[V. Enss \and V. Kostrykin \and R. Schrader]{Volker Enss \and
Vadim Kostrykin
\and Robert Schrader$^{\ast}$}

\address{Volker Enss\\ Institut f\"{u}r Reine und Angewandte Mathematik\\
Rheinisch-Westf\"{a}lische Technische Hochschule Aachen,
Templergraben 55, D-52062 Aachen,
Germany}
\email{enss@rwth-aachen.de}

\address{Vadim Kostrykin\\
Fraunhofer-Institut f\"{u}r Lasertechnik\\ Steinbachstra{\ss}e 15,
D-52074\\ Aachen, Germany}
\email{kostrykin@t-online.de, kostrykin@ilt.fraunhofer.de}

\address{Robert Schrader\\ Institut f\"{u}r
Theoretische Physik\\ Freie Universit\"{a}t Berlin, Arnimallee 14\\
D-14195 Berlin, Germany}
\email{schrader@physik.fu-berlin.de}

\thanks{$^\ast$ R.S. supported in part by
DFG SFB 288 ``Differentialgeometrie und Quantenphysik''}



\begin{document}

\begin{abstract}
Quantum mechanical scattering theory is studied for time-dependent Schr\"{o}dinger operators, in
particular for particles in a rotating potential. Under various assumptions about the decay rate
at infinity we show uniform boundedness in time for the kinetic energy of scattering states,
existence and completeness of wave operators, and existence of a conserved quantity under
scattering. In a simple model we determine the energy transfered to a particle by a collision with
a rotating blade.
\end{abstract}

\begin{flushleft}
\textsf{\tiny Proceedings of the Workshop on}\\
\vspace{-1mm}
\textsf{\tiny Spectral and Inverse Spectral Problems for Schr\"{o}dinger
Operators}\\
\vspace{-1mm}
\textsf{\tiny Goa, India, December 14--20, 2000}\\
\vspace{5mm}
\end{flushleft}

\maketitle

\section{Introduction}

This note is a preliminary report on the study of explicitly time-dependent
periodic Schr\"{o}dinger operators on $L^2(\R^\nu)$, $\nu\geq 2$
\begin{equation}\label{Ham}
H(t)= H_0 + V_t\, ,\quad H_0 =  -\frac{1}{2m}\,\Delta
\end{equation}
with ``rotating" potentials of the form
\begin{equation}\label{2:alt}
V_t(\x) = V(\cR(t)^{-1} \x),
\end{equation}
where $V$ is some time-independent function decaying at infinity and $\cR(t)$ is
a rotation by $\omega t$ in the $x_1,x_2$-plane with period $2\pi/\omega$,
\begin{equation*}
\begin{split}
(\cR(t) \x)_1 &= \cos(\omega t) x_1 - \sin(\omega t) x_2,\\ (\cR(t)
\x)_2 &=
\sin(\omega t) x_1 + \cos(\omega t) x_2,\\
(\cR(t) \x)_k &= x_k,\qquad k=3,\ldots,\nu.
\end{split}
\end{equation*}

An important difference between time-independent and time-dependent perturbations is
that the latter do not conserve the energy. See, e.g., \cite{Howland:79, Yajima:77,
Kitada:Yajima} for studies of time periodic potentials. If one knows that the kinetic
energy remains uniformly bounded (or increases at most logarithmically in time) then
(cf.\ \cite{Enss:Veselic}) the machinery of time-dependent scattering theory
\cite{Enss} applies giving the existence and completeness of the wave operators. Some
sufficient conditions for the boundedness of energy (in the sense of definition on
p.\ 171 of \cite{Enss:Veselic}, see \eqref{2.7A}) for repulsive potentials (in
particular, of the form \eqref{2:alt}) are given by Huang and Lavine
\cite{Huang:Lavine, Huang}, for smooth potentials e.g.\ by Nakamura \cite{Nakamura}.

In the context of classical mechanics a somewhat similar question of energy transfer and
boundedness was recently discussed in a study of dynamics of black holes \cite{Hawking}. We
mention also the work of Cooper and Strauss (see \cite{Cooper:Strauss} and references therein)
where the scattering off periodically moving obstacles and the boundedness of energy for
scattering states have been considered for the wave equation in the framework of Lax-Phillips
theory.

In Section~2 we show boundedness of the kinetic energy on the ranges of wave operators for a wide
class of potentials. In Sections~3 and 4 we study the time evolution in a
rotating frame for potentials which need not be smooth. This transformation which has a well-known
counterpart in classical mechanics (see, e.g., Example 2 in Section 5.33 of
\cite{Dubrovin:Novikov:Fomenko} or Section 39 of \cite{Landau:Lifshits}) yields an explicit
formula for the propagator $U(t,s)$. Methods of stationary scattering theory can then be applied
to show existence and completeness of the wave operators. In the final section we discuss a simple
model which describes the energy transfer between a quantum particle and a rotating blade.

\section{Boundedness of Kinetic Energy}
\setcounter{equation}{0}

In this section we study bounds of the kinetic energy on incoming and outgoing
scattering states. These bounds follow from suitable decay assumptions on the
potential. If one knows that the scattering operator $S$ is unitary (i.e., $\Ran\
\Omega^+ = \Ran\ \Omega^-$) or even that scattering is asymptotically complete then
we will show that the kinetic energy is bounded uniformly in both time directions on
all scattering states.

We begin with a rather abstract proposition. Then we show that certain
classes of potentials satisfy the assumptions in the proposition. The concrete
form of $H_0$ does not matter,
\begin{equation*}
H_0 = \frac{1}{2m} p^2 = - \frac{1}{2m} \Delta,\qquad H_0=\sqrt{p^2 c^2 + m^2
c^4},\qquad  \ldots
\end{equation*}
or Dirac operators (with some straightforward modifications) can be treated equally
well. It is only the propagation properties in configuration space under the free
time evolution analogous to \eqref{14} and \eqref{15} which matter in the
applications. Throughout this section we assume for simplicity of presentation that
the potentials $V_t$ (which need not be of the special form \eqref{2:alt}\,) are
uniformly Kato-bounded with respect to a free Hamiltonian $H_0$, i.e., there are
constants $a<1$ and $b<\infty$ such that
\begin{eqnarray}
&& \|V_t \Psi\| \leq a \| H_0 \Psi\| + b \|\Psi\|\qquad
\forall\Psi\in\cD(H_0),\quad t\in\R\, ,
\label{1}\\
&& \sup_{t\, \in\R} \|\partial_t V_t\| < \infty.\label{2}
\end{eqnarray}
In particular, $H(t) = H_0 + V_t$ is self-adjoint on $\cD(H(t)) =
\cD(H_0)$ for all $t\in\R\,$.

\begin{proposition}\label{erste:Proposition}
Let $H(t) = H_0 + V_t$ be a self-adjoint family of operators which satisfies
\eqref{1}, \eqref{2} and generates a unitary propagator $U(t,s)$ with
$U(t,s)\; \cD(H_0) \subseteq \cD(H_0)$ and
\begin{equation}\label{3}
U(s,s) = I,\qquad i\frac{d}{dt} U(t,s)\;\Psi_s = H(t) \;U(t,s)\;\Psi_s
\end{equation}
for $\Psi_s\in\cD(H_0)$. Let the perturbation $V_t$ satisfy the following
conditions: There is a total set $\cD_0$ such that for any $\Phi\in\cD_0$ there
is a positive integrable function $h\in L^1(\R)$ (depending on $\Phi$) with
\begin{eqnarray}
&& \| V_t\; e^{-i H_0 t}\; \Phi \| \leq \frac{h(t)}{1+|t|},\label{5}\\ && \|
\partial_t V_t\; e^{-i H_0 t}\; \Phi \| \leq h(t).\label{6}
\end{eqnarray}
Then the wave operators
\begin{equation}\label{4}
\Omega^\pm = \slim_{t\rightarrow\pm\infty} U(t,0)^\ast\; e^{-iH_0 t}
\end{equation}
exist and the kinetic energy is uniformly bounded in time on the ranges of
$\Omega^\pm$ in the following sense: For $\Psi\in\Omega^\pm\cD_0$, a total set
in $\Ran\ \Omega^\pm$,
\begin{equation}\label{7}
\sup_{\pm\, t\, \geq\ 0} \| H_0^{1/2}\; U(t,0)\; \Psi\| \leq
\mathrm{const}.
\end{equation}
This implies for every $\epsilon > 0$ and $\Psi \in \Ran\
\Omega^\pm$ that there exists a cutoff energy $E(\epsilon, \Psi)$
such that
\begin{equation}\label{2.7A}
\sup_{\pm\, t\, \geq\ 0} \| F(H_0 > E(\epsilon, \Psi)\,)\;\; U(t,0)\; \Psi\|
< \epsilon.
\end{equation}
\end{proposition}

\textbf{Remarks.} The condition of boundedness \eqref{2} can be replaced by
much weaker conditions of relative boundedness in the case of
specific time evolutions like \eqref{equality} for rotating potentials.
The conditions \eqref{5} and \eqref{6} alone do not guarantee the existence
of the unitary propagator satisfying \eqref{3}; very general sufficient
conditions are given in \cite{Yajima:87}.
\begin{proof}
The wave operators $\Omega^\pm$ exist since the norm in \eqref{5}
is integrable with respect to $t$. They are unitary as maps
$\Omega^\pm\ :\ \cH \rightarrow \Ran\ \Omega^\pm$. For any total
set $\cD_0\subset\cH$ the images $\Omega^\pm \cD_0$ are total in
$\Ran\ \Omega^\pm$.

The uniform Kato boundedness of the potentials implies a form bound
\begin{equation*}
|(\Psi,\; V_t \Psi)| \leq a' (\Psi,\; H_0 \Psi) + b' \|\Psi\|^2
\end{equation*}
with $a'<1$ for all $\Psi\in\cD(H_0)$ and $t\in\R$. Thus, for
$\Psi\in\cD(H_0)$ we can use the obvious estimate
\begin{equation*}
(\Psi,\; H_0 \Psi) \leq
\frac{1}{1-a'}\left\{|(\Psi,\; H(t) \Psi)| + b'\|\Psi\|^2
\right\}.
\end{equation*}
To verify \eqref{7} it is sufficient to show a bound for
\begin{equation}\label{9}
\sup_{\pm\ t\ \geq\ 0} \left|\left(U(t,0)\;\Psi,\; H(t)\; U(t,0)\;\Psi
\right)\right|
\end{equation}
for suitable $\Psi = \Omega^\pm\; \Phi \in \cD(H_0)$. The time derivative
of the scalar product exists and is of the form
\begin{equation}\label{10}
\frac{d}{dt} (U(t,0)\;\Psi,\; H(t)\; U(t,0)\;\Psi) = (U(t,0)\;\Psi,\;
\partial_t V_t\; U(t,0)\;\Psi).
\end{equation}
The supremum in \eqref{9} is finite if
\begin{equation*}
\|\partial_t V_t\; U(t,0)\;\Psi\| \leq
\|\partial_t V_t\|\ \|U(t,0)\;\Psi-e^{-i H_0 t}\;\Phi\| +
\|\partial_t V_t\; e^{-i H_0 t}\;\Phi\|
\end{equation*}
is integrable on $\pm\ t \in [0,\infty)$. By assumption \eqref{6}
this follows for the second term on the r.h.s.\ for a total set of
states $\Psi$.

For $\Psi = \Omega^\pm\; \Phi$ we have
$\displaystyle\lim_{t\rightarrow +\infty}
U(t,0)^\ast\; e^{-i H_0 t}\; \Phi = \Psi$. Thus
\begin{equation*}
\begin{split}
\|U(t,0)\;\Psi-e^{-i H_0 t}\;\Phi\| &=
\|\Psi - U(t,0)^\ast\; e^{-i H_0 t}\;\Phi\|\\
& \leq \int_s^\infty ds \|V_s\; e^{-i H_0 s}\; \Phi\| = \int_t^\infty ds
\frac{h(s)}{1+|s|}
\end{split}
\end{equation*}
for some integrable function $h\in L^1([0, +\infty))$ by assumption \eqref{5}.
Using partial integration we conclude integrability:
\begin{equation*}
\int_0^\infty dt\int_t^\infty ds \frac{h(s)}{1+s} =
t\int_t^\infty ds \frac{h(s)}{1+s}\Big|_{t=0}^{t=\infty} + \int_0^\infty dt
\frac{t}{1+t} h(t) \leq \int_0^\infty ds h(s) < \infty.
\end{equation*}
Consequently, the time derivative \eqref{10} is integrable on $[0,\infty)$ and
the supremum \eqref{9} is finite for a total set of $\Psi = \Omega^+\; \Phi$,
$t\geq 0$. The uniform boundedness for $t\leq 0$ and $\Psi = \Omega^-\; \Phi$
is proved similarly.
\end{proof}

\vspace{5mm}

Next we will give sufficient conditions which guarantee that
\eqref{5} and \eqref{6} are satisfied. For simplicity of
presentation we use standard nonrelativistic kinematics \eqref{Ham}, $H_0 =
p^2/2m$. We will apply geometrical time-dependent methods. Then a convenient
total set $\cD_0\subset\cH$ consists of states with good localization in
momentum space. Let $\widehat{\varphi}(\p)$ denote the momentum space wave
function of $\Phi$ and $B_{m \vitsmall/3}(m
\v)\subset\R^\nu$ the open ball of radius $m \vit/3$ with
center $m \v\in\R^\nu$, $\v\neq 0$, $\vit = |\v|$. We choose the set $\cD_0$ as
\begin{equation}\label{13}
\cD_0\ :=\ \{\Phi\in\cH\mid \|\Phi\|=1,\ \widehat{\varphi}\in
C_0^\infty(\R^\nu),\ \exists \v\in\R^\nu\ \text{such that}\ \supp\
\widehat{\varphi}\subseteq B_{m \vitsmall/3}(m
\v)\}.
\end{equation}
Any state $\Psi$ with $\widehat{\psi}\in C_0^\infty(\R^\nu)$, $0\notin \supp\
\widehat{\psi}$ can be written as a finite linear combination of vectors in $\cD_0$.
This set is dense in $L^2(\R^\nu) = \cH$.

The states in $\cD_0$ propagate mainly into regions where
$\x\approx t \p /m\approx t\v$, $\p\in\supp\ \widehat{\varphi}$. More
precisely, one shows with a stationary phase estimate that
propagation into ``classically forbidden'' regions decays rapidly:
\begin{equation}\label{14}
\| F(|\x - t\v| \geq \rho + |t|\vit/2)\;e^{-itH_0}\;\Phi \|
\leq C_N\;(1+\rho + |t|)^{-N},\;
N\in\N,\; \rho\geq 0,
\end{equation}
with a constant $C_N=C_N(\Phi) < \infty$ (see, e.g., Section II of \cite{Enss:83}).
Similar estimates hold for other kinematics. We will use this bound for $\rho=0$ here
and with $\rho>0$ in the last section.

While the estimate \eqref{14} follows from propagation of wave packets one has,
in addition, the standard estimate of spreading in $\R^\nu$
\begin{equation}\label{15}
\sup_{x\, \in\R^\nu} |(e^{-i H_0 t}\ \Phi)(x)|\quad \leq\quad
C(\Phi)\; (1+|t|)^{-\nu/2},
\end{equation}
where $C(\Phi)<\infty$ for $\Phi\in\cD_0$.

Now we return to rotating potentials \eqref{2:alt} which are
possible in $\nu\geq 2$ dimensions. We will give sufficient
conditions for the two dimensional case which is the ``worst case": the
falloff \eqref{15} is slowest and -- compared to $\R^3$ -- the
potential does not decay in the direction parallel to the axis of
rotation. We may use polar coordinates $(r,\phi)$ in the $(x_1,
x_2)$-plane.

The potential can be decomposed into a rotationally invariant part
\begin{equation*}
V_{\mathrm{inv}}(x)\quad :=\quad \frac{\omega}{2\pi}
\int_0^{2\pi/\omega} V(\cR(t)^{-1}x)\ dt
\end{equation*}
and the rest $V_{\mathrm{noninv}} = V - V_{\mathrm{inv}}$. The rotationally invariant
part of the potential remains time-independent. It need not be bounded nor
differentiable and it does not show up in \eqref{6}. If for every $g\in
C_0^\infty(\R)$ there is an integrable $\widetilde{h}\in L^1([0,\infty))$ (e.g.,
$\widetilde{h}(\rho) = C(1+\rho)^{-1-\epsilon}$) such that
\begin{equation}\label{16}
\| V_{\mathrm{inv}}\ g(H_0)\ F(|x|>\rho)\|\quad \leq \quad
\frac{\widetilde{h}(\rho)}{1+\rho}
\end{equation}
then \eqref{5} is satisfied for $V_{\mathrm{inv}}$: For
$\Phi\in\cD_0$ choose $g\in C_0^\infty(\R)$ such that $g(H_0)\Phi
= \Phi$. Then
\begin{equation*}
\begin{split}
\|V_{\mathrm{inv}}\ e^{-i H_0 t}\ g(H_0)\ \Phi\|\quad \leq\quad &
\|V_{\mathrm{inv}}\ g(H_0)\ F(|x|>|t|\vit / 2)\|\ \|\Phi\| \\
&+\|V_{\mathrm{inv}}\ g(H_0)\|\
\|F(|x|<|t|\vit / 2)\ e^{-i H_0 t} \Phi\|\\
\leq\quad & \frac{\widetilde{h}(|t|\vit / 2)}{1+|t|\vit  / 2} +
O(|t|^{-N}) = \frac{h(t)}{1+|t|}
\end{split}
\end{equation*}
with $h\in L^1$ by \eqref{16} and \eqref{14}.

\begin{lemma}\label{lem2.2}
Let $V$ be Kato-bounded and let there exist an integrable function
$h \in L^1([0,\infty))$ such that the potential $V$ satisfies the condition
\begin{equation}\label{18}
\rho\;\| V\;F(|\x| > \rho)\| \leq h(\rho)
\end{equation}
or one of the weaker conditions
\begin{equation}\label{19}
\rho\;\| V\;(H_0 + 1)^{-1}\;F(|\x| > \rho)\| \leq h(\rho)
\end{equation}
or for every $g\in C_0^\infty (\R)$ there is an integrable $h = h_g$ with
\begin{equation}\label{20}
\rho\;\| V\;g(H_0)\;F(|\x| > \rho)\| \leq h(\rho).
\end{equation}
Then the rotating potential $V_t = V(\cR(t)^{-1} \cdot)$ satisfies
\eqref{5}, i.e., for every $\Phi \in \cD_0$ \eqref{13} there is an
integrable $\widetilde{h}$ such that
\begin{equation*}
|t|\; \|V_t\; e^{-itH_0}\; \Phi\| \leq \widetilde{h}(|t|).
\end{equation*}
If the partial (distributional) azimuthal derivative $(\partial_\phi V)(r,\phi)$
yields a bounded multiplication operator $ \partial_\phi V$ which satisfies
\begin{equation}\label{21}
\|\partial_\phi V\;F(|\x|>\rho)\| \leq h(\rho)
\end{equation}
or the weaker
\begin{equation}\label{22}
\|\partial_\phi V\;(H_0 +1)^{-1}\;F(|\x|>\rho)\| \leq h(\rho)
\end{equation}
or for every $g\in C_0^\infty(\R)$
\begin{equation}\label{23}
\|\partial_\phi V\;g(H_0)\; F(|\x|>\rho)\| \leq h(\rho)
\end{equation}
for some integrable $h$ then \eqref{6} holds, i.e., for every $\Phi\in\cD_0$ there is
an integrable $\widetilde{h}$ with
\begin{equation*}
\|\partial_t V_t\; e^{-itH_0}\; \Phi\| \leq \widetilde{h}(|t|).
\end{equation*}
\end{lemma}
\noindent\textbf{Remarks.} If \eqref{18} holds then it implies
\eqref{19} and \eqref{20} because the regularizing factors $(H_0 +
1)^{-1}$ or $g(H_0)$ act in configuration space as convolutions
with a continuous rapidly decreasing function. Thus the required
decay rate is preserved. But even if the operators on the l.h.s.\
of \eqref{18} are bounded the decay rate may be better in the
regularized versions \eqref{19} or \eqref{20}: think of a sequence
of ``dipole'' pairs of peaks with maxima and minima of equal
amplitude but ``closer and thinner'' pairs when they are localized
farther away. Then $\|V\; F(|\x|>\rho)\|$ does not decay but the
convolution causes falloff due to cancellations. The same applies to
conditions \eqref{21} -- \eqref{23}.

A potential $V(r,\phi)$ which in an angular sector behaves like
\begin{equation*}
V(r,\phi) = \frac{1}{r^2\;(\ln r)^2}\,\cos(r^\alpha\,\phi),\; r>2,\;
\phi_1<\phi<\phi_2,
\end{equation*}
satisfies there \eqref{18} and \eqref{21} for exponents $0\leq\alpha \leq 1$
but the latter is violated for $\alpha >1$. A behavior like $\alpha =1$ will
show up in the next example.
\begin{proof}[Proof of Lemma~\ref{lem2.2}]
Since $\Phi \in \cD_0$ has compact support in momentum space we may choose
$g\in C_0^\infty(\R)$ such that $g(H_0)\;\Phi = \Phi$. Due to rotational
invariance of $H_0$ and $|\x|$ we have
\begin{equation*}
\begin{split}
\| V_t\; g(H_0)\; F(|\x|>\rho)\|\ & =\ \| V\; g(H_0)\; F(|\x|>\rho)\|,\\
\|\partial_t V_t\; g(H_0)\; F(|\x|>\rho)\|\ & =\
\omega\; \|\partial_\phi V\; g(H_0)\; F(|\x|>\rho)\|.
\end{split}
\end{equation*}
To estimate \eqref{5} we use \eqref{20} and \eqref{14}:
\begin{align*}
\| V_t\;e^{-itH_0}\,\Phi \|&\leq \| V\;g(H_0)\;F(|\x| > |t|\,\vit/2)\|\;\|\Phi \|
\notag \\
&\quad + \| V\;g(H_0)\| \;\| F(|\x| < |t|\,\vit/2)\;e^{-itH_0}\,\Phi \| \notag
\\ &\leq \frac{1}{1+|t|\,\vit/2}\,h(|t|\,\vit/2) + O(|t|^{-N}).
\end{align*}
Similarly, \eqref{23} and \eqref{14} yield \eqref{6}.

In the case of regularization with a resolvent observe that
$(H_0+1)^{-1}\;\Phi / \| (H_0+1)^{-1}\;\Phi \| \in \cD_0$ has the
same smoothness and support properties in momentum space as
$\Phi$.
\end{proof}
Another geometrical configuration is described by a strongly anisotropic potential localized near
a hyperplane, in $\nu=2$ dimensions near a
line. For simplicity we assume that the support is bounded in the $x_2$-direction, a sufficiently
rapid decay would give the same result. Moreover, we state the lemma for differentiable potentials
in product form, the generalization to less regular ones as in the previous lemma is
straightforward.

\begin{lemma}
Let the potential $V(x_1, x_2) = V^{(1)}(x_1)\, V^{(2)}(x_2) \in
C^1(\R^2)$ satisfy $\supp\  V^{(2)} \subset [-d,d]$ and the bound
\begin{equation}\label{28}
\rho^{1/2}\;\sup_{|x_1|\geq \rho} \left| V^{(1)}(x_1)\right| \quad +\quad
\left(\frac{1}{1+\rho}\right)^{\!\! 1/2}\sup_{|x_1|\geq \rho} \left|
\frac{d}{dx_1}V^{(1)}(x_1)\right| \quad\leq\; h(\rho)
\end{equation}
for some integrable $h$. Then $V_t= V(\cR(t)^{-1} \cdot)$ satisfies conditions
\eqref{5} and \eqref{6} for every $\Phi\in\cD_0$.
\end{lemma}
\begin{proof}
Up to rapidly decaying parts which do not affect the integrability the
configuration space wave function is localized in a moving disk and satisfies
for large $|t|$ the estimate
\begin{equation*}
\left|\left(e^{-itH_0}\,\Phi\right)\!(\x)\right| \leq
\frac{\mathrm{const}}{|t|}\;\chi_{B_{|t|\, \vitsmall/2}(t \v)}(\x)
\end{equation*}
by \eqref{14} and \eqref{15}. $\chi_{B_{|t|\, \vitsmall/2}(t \v)}$ denotes the
characteristic function of $B_{|t|\, \vitsmall /2}(t \v)$. The $k$-th passage
of a ``tail'' of the rotating potential takes place around $t_k=k\pi/\omega$
and lasts less than $2\tau=\pi/\omega$ (for $|t|>5d/\vit$). The area of
intersection of the disk with the support of the potential is bounded by $d
\vit\, (|t_k| +
\tau)$ and
\begin{equation*}
|V(\x)| \leq \sup\ |V^{(2)}|\;\; \frac{1}{\vit\, (|t_k| - \tau)
/2}\,h(\vit\, (|t_k| - \tau) /2),\quad \x\in B_{|t|\, \vitsmall/2}(t \v),\;|t|
\geq |t_k|-\tau .
\end{equation*}
For given $\v$ and $\omega$ we obtain for one passage (up to rapidly decaying
terms)
\begin{align}\label{31}
&\int_{t_k-\tau}^{t_k+\tau} dt\;\left\|V_t\,e^{-itH_0}\,\Phi\right\| \notag \\
&\leq 2\tau\;\frac{1}{\vit\, (|t_k| + \tau) /2}\,h(\vit\, (|t_k| + \tau)
/2)\;\;
\frac{\mathrm{const}}{|t_k| - \tau}\;\left\{ d \vit (|t_k| + \tau)\right\}^{1/2}
\notag \\[0.5ex]
& \leq \frac{\mathrm{const}}{|t_k| + \tau}\;h({\rm const}\,|t_k|)
\end{align}
for large enough $|k|$. Since $\|V_t\,e^{-itH_0}\,\Phi\|$ is bounded on compact
intervals the estimate \eqref{31} shows \eqref{5}.

With $\partial_t V_t\ =\ \omega\, [\,x_2\partial_1 V - x_1 \partial_2 V\,]
(\cR(t)^{-1}\cdot)$ the first summand yields a bound on $B_{|t|\, \vitsmall/2}(t
\v)$, $|t|\geq |t_k|-\tau$
\begin{equation*}
\omega\;\sup_{x_2}\left|x_2\,V^{(2)}(x_2)\right|\;
[\, \vit\, (|t_k| - \tau) /2 \,]^{1/2}\; h(\vit\, (|t_k| - \tau) /2)
\end{equation*}
by \eqref{28} while the second is bounded there by
\begin{equation*}
\omega\;\sup_{x_2}\left|\frac{d}{dx_2}\,V^{(2)}(x_2)\right|\;
\frac{3\vit\, (|t_k| - \tau) /2}{[\vit\, (|t_k| - \tau) /2]^{1/2}}\;
h(\vit\, (|t_k| - \tau) /2).
\end{equation*}
Combining these estimates as above shows \eqref{6}.
\end{proof}

Our third example demonstrates how dimensions strictly larger than two help
if the potential decays in the other directions. For simplicity we
assume $\nu=3$ and compact support in the vertical direction
(parallel to the axis of rotation) of a differentiable potential.
Note that we do not need any falloff in the plane of rotation to
show boundedness of the kinetic energy for asymptotically free
scattering states. (The existence of wave operators follows easily
for such potentials but one will need additional assumptions for
asymptotic completeness.)

\begin{lemma}\label{lem2.3}
Let $V\in C^1(\R^3)$ have bounded $C^1$-norm and satisfy $\supp\ V
\subset \{ \x\in\R^3 \mid |x_3|\leq d \}$. Then \eqref{5} and
\eqref{6} are satisfied.
\end{lemma}
\begin{proof}
Let $\cD_0$ be the total set of states with $\widehat{\varphi}\in
C_0^\infty (\R^3)$, for which there exists a constant $b>0$ such
that either $\supp\ \widehat{\varphi} \subset \{\p\in\R^3 \mid p_3 >
mb\}$ or $\supp\ \widehat{\varphi} \subset \{\p\in\R^3 \mid p_3 <
-mb\}$. Then $\|F(|x_3|< |t|b/2)\;e^{-itH_0}\;\Phi\| =
O(|t|^{-N})$ and conditions \eqref{5} and \eqref{6} follow.
\end{proof}

To sum up the results of this section: If one knows (using any method)
unitarity of the scattering operator or even asymptotic completeness and if the
potential can be split into a sum of terms which satisfy any of the above
sufficient conditions, then the kinetic energy is bounded uniformly in time
in both time-directions simultaneously on the corresponding subspace of
asymptotically free scattering states.

\section{Evolution in a Rotating Frame}
\setcounter{equation}{0}

Here we study the time evolution in a rotating frame for potentials which no
longer have to be smooth. This
transformation yields an explicit formula for the propagator $U(t,s)$ in terms of the unitary
group for some time-independent generator. This will allow to apply
methods of stationary
scattering theory to show existence and completeness of the wave operators
in Section~4.

Let $\cR(t) \mapsto R(t)$ be the standard unitary representation of the one-parameter
group $\cR(t)$ in $L^2(\R^\nu)$, i.e., $(R(t)\psi)(x) =
\psi(\cR(t)^{-1} x)$. Let $\omega J$ denote its generator, $R(t) =
\exp\{-i \omega t J \}$. On a suitable domain the operator $J$ is
of the form $x_1 (-i \partial/\partial x_2) - x_2(-i\partial/\partial x_1)$ or
$-i\partial/ \partial \phi$ if one uses Cartesian or polar coordinates,
respectively, in  the $x_1, x_2$-plane.

For an observer in a rotating reference frame which turns around the orgin
like the potential the latter becomes time-independent
\begin{equation*}
V_t = R(t)\; V \;R(t)^* \longrightarrow R(t)^* \;V_t\;R(t) = V.
\end{equation*}
Let $\, t\mapsto \Psi(t) = U_{\mathrm{inert}}(t,s)\,\Psi(s)$ be \emph{any}
time evolution in the given inertial frame with propagator
$U_{\mathrm{inert}}$. Then an observer in the rotating frame will see
\begin{equation*}
R(t)^*\,\Psi(t) = R(t)^*\,U_{\mathrm{inert}}(t,s)\,\Psi(s) =
R(t)^*\,U_{\mathrm{inert}}(t,s)\,R(s)\; R(s)^*\,\Psi(s)
\end{equation*}
with propagator
\begin{equation}\label{rotframe}
U_{\mathrm{rot}}(t,s) =  R(t)^*\,U_{\mathrm{inert}}(t,s)\,R(s).
\end{equation}

The free time evolution of a state then becomes
\begin{equation}\label{e3.1}
R(t)^* \;e^{-itH_0}\,\Psi = e^{it \omega J}\;e^{-itH_0}\,\Psi
\end{equation}
where $e^{-itH_0}\,\Psi$ is the free time evolution in the inertial frame
generated by $H_0$ as in \eqref{Ham} (or any other spherical free
Hamiltonian like the relativistic one). Time
zero (or $k\,2\pi/\omega$, $k\in\Z$) is singled out by the fact that the
rotating and inertial frames coincide and the fixed potential $V_t|_{t=0} = V$
has been picked out of the family $V_t$ for this reference time. Although the
free time evolution is rotation invariant we have a different ``unperturbed''
evolution which combines the unchanged free evolution with the rotation.
Instead of a motion with constant velocity the unperturbed motion now is
along spirals.

As the groups in \eqref{e3.1} commute their product is again a unitary group
with a self-adjoint generator denoted by $H_\omega$
\begin{equation*}
e^{i\omega J}\;e^{-itH_0} =: e^{-it H_\omega} .
\end{equation*}
Formally we have
\begin{equation}\label{e3.3}
H_\omega =H_0 - \omega J
\end{equation}
but the domains differ. All three operators are essentially self-adjoint on
each of the sets
\begin{equation}\label{e3.4}
\cD:= \{\Psi \in \cH \mid \widehat\psi \in C_0^\infty (\R^\nu )\}
\subset \cS(\R^\nu)\subset \cD(H_0) \cap \cD(J)
\end{equation}
where $\cD$ is the set of states with smooth compactly supported wave functions in
momentum space, $\cS(\R^\nu)$ the Schwartz space of smooth rapidly decreasing
functions (in configuration or momentum space) and $\cD(A)$ denotes the domain of a
self-adjoint operator $A$. All these sets are cores because they are dense in
$L^2(\R^\nu)$ and invariant under each of the groups (see, e.g.,
\cite[Theorem~VIII.11]{RS1}).

The operator \eqref{e3.3} has been previously studied by Tip \cite{Tip} in
connection with the
circular AC Stark effect. Let $P_j$, $j\in\Z$ denote the projection onto the
eigenspace of $J$. Since $H_0$ and $J$ commute, the subspaces $\cH_j = P_j
\cH$ are invariant subspaces for $H_\omega$ such that
\begin{equation*}
H_\omega=\bigoplus_{j\in\Z} H_{\omega,j} = \bigoplus_{j\in\Z}
\left(H_{0j}- \omega j\right).
\end{equation*}
In the momentum representation $H_{0j}$ is a real multiplication operator and
consequently $H_{\omega,j}$ with domain
$\cD_j=(H_{\omega,j}-i)^{-1}\cH_j\subset\cH_j$ is self-adjoint on $\cH_j$.
Let now
\begin{equation*}
\cD(H_\omega):=\left\{f=\bigoplus_j f_j \biggm| f_j\in\cD_j,\
\sum_j \|H_{\omega,j}\: f_j\|_j^2 < \infty \right\}
\end{equation*}
with $\|\cdot\|_j$ being the norm in $\cH_j$. The operator $H_\omega$ with
the domain $\cD(H_\omega)$ can be easily shown to be self-adjoint.
Its domain is rotational invariant $R(t)\,\cD(H_\omega) =
\cD(H_\omega)$ and the operator commutes with rotations.

The set $\cD(H_\omega)$ is strictly larger than $\cD(H_0)\cap
\cD(J)$. Indeed, consider a state $\Psi_0\in\cH$ with $\|\Psi_0\|=1$
which in the momentum representation is given by the function
$\widehat{\psi}_0\in C_0^\infty$. We assume that
\begin{equation*}
\supp\ \widehat{\psi}_0 \subset\{\p\in\R^\nu\mid
|\p|<1/2\}
\end{equation*}
and $\widehat{\psi}_0(\p)$ is rotational symmetric
such that $\int p_1\,|\widehat{\psi}_0(\p)|^2\,dp =0$.
For $n\in\N$ and $\omega\neq 0$ consider the sequence of normalized pairwise
orthogonal vectors in $\cD$ \eqref{e3.4}
\begin{equation*}
\widehat{\psi}_{\omega,n}(\p) := \exp\left\{i n\frac{ p_2}{2m\,\omega}\right\}
\widehat{\psi}_0(\p-n \e_1)
\end{equation*}
with $\e_1$ being the unit vector in $p_1$ direction. These states are
essentially localized in momentum space near $n\e_1$ and in configuration
space near $(n/2m\omega)\e_2$. Simple calculations give
\begin{equation*}
\|2m\,H_0 \Psi_{\omega,n}\|^2 = \int_{\R^\nu} |\p|^4 \,
|\widehat{\psi}_0(\p- n \e_1)|^2 \, dp
= \int_{\R^\nu} |\p+n \e_1|^4\, |\widehat{\psi}_0(\p)|^2 \, dp
= n^4 + O(n^2)
\end{equation*}
because the term proportional to $n^3$ vanishes by symmetry.
Further we estimate the norm of $\omega J\,\Psi_{\omega,n}$. In the momentum
representation we have
\begin{equation*}
(\omega J\,\widehat{\psi}_{\omega,n})(\p) =
i\omega \frac{\partial}{\partial p_1}(p_2
\widehat{\psi}_{\omega,n})(\p)) - i\omega  \frac{\partial}{\partial p_2}(p_1
\widehat{\psi}_{\omega,n})(\p)).
\end{equation*}
The first term is obviously bounded uniformly in $n$. The second term can
be written in the form
\begin{equation}\label{e3.6}
\frac{n}{2m}\exp\left\{in\frac{p_2}{2m\,\omega}\right\}
p_1 \widehat{\psi}_0(\p- n \e_1)
-i\omega  \exp\left\{in\frac{ p_2}{2m\,\omega}\right\}
p_1 \frac{\partial}{\partial p_2} \widehat{\psi}_0(\p- n \e_1).
\end{equation}
For large $n$ the first summand is the dominant
contribution. Again, the square of the norm of \eqref{e3.6} is
$(n^2 /2m)^2 + O(n^2)$. Now we turn to the estimate of
$\|H_\omega\Psi_{\omega, n}\|$.
The leading terms cancel in
\begin{equation*}
\left(\frac{p_1}{2m} + i\omega \frac{\partial}{\partial p_2} \right)
p_1\widehat{\psi}_{\omega, n}(\p).
\end{equation*}
and one obtains easily that $\|H_\omega\Psi_{\omega, n}\|^2 =O(n^2)$ or better.

Thus, we have shown that for large $n$ the norms $\| H_0\Psi_{\omega, n}\|$ and
$\|J\,\Psi_{\omega, n}\|$ are of order of magnitude $O(n^2)$ whereas the norm
$\|H_\omega\Psi_{\omega, n}\|$ is of order of magnitude $O(n)$. Choose an
arbitrary sequence of coefficients $\{\alpha_n\}_{n\in\N_0}$ such that
$\sum_{n=0}^\infty n^2 |\alpha_n|^2<\infty$ but $\sum_{n=0}^\infty n^4
|\alpha_n|^2$ diverges. Let $\widehat{\Psi}=\sum_n
\alpha_n \widehat{\Psi}_{\omega,n}$; by the preceding estimates it is contained
in $\cD(H_\omega)$ but
neither in $\cD(H_0)$ nor $\cD(J)$. Thus $J \,(H_\omega-i)^{-1}$
is not a bounded operator! This means that there are quantum states for which
the quantity $H_0 - \omega J$ is bounded but both the angular momentum and the
kinetic energy are unbounded.

A similar calculation shows the corresponding statement for quadratic forms. $\widehat{\Psi} \in
\cQ (H_\omega)$, the form domain, for any square summable sequence of coefficients but
$\widehat{\Psi} \notin \cQ (H_0)$ and $\widehat{\Psi} \notin \cQ (J)$ as soon as
$\sum_{n=0}^\infty n^2 |\alpha_n|^2$ diverges.  This can happen, however, only for states with a
bad localization in configuration and momentum space and a good correlation like $(p_1/2m)\sim
\omega x_2$. In particular, the domains of self-adjointness of $H_\omega \;``= H_0 - \omega J$''
are pairwise different for different values of $\omega$. A further technical complication is the
fact that $H_\omega$ is not bounded below.

For $\alpha\in\R\,$ we define
\begin{equation}
G^\alpha (\x) =(1+|\x|^2)^{\alpha}, \qquad G^\alpha\;G^{-\alpha} =1;\qquad
\| G^\alpha\| =1 \;\;\text{ if }\;\; \alpha \leq 0.
\end{equation}
We will need the following lemma, which is a variant of a results of
Tip \cite[Lemma~2.1 and 2.2]{Tip}:

\begin{lemma}\label{lemma:Tip}
Let $\Phi \in \cS$, the Schwartz space of rapidly decreasing functions. Then
for all $\alpha \geq 1$ $G^{-\alpha}
\Phi\in\cD(H_0)$ and for any arbitrarily small $\epsilon>0$
\begin{equation*}
\| H_0\; G^{-\alpha}\; \Phi\| \leq (1+\epsilon)\ \|H_\omega\; \Phi\| +
b(\epsilon)\ \|\Phi\|,
\end{equation*}
with $b(\epsilon)$ being non-negative.

If $(1+|\x|^2)\;V$ is bounded relative to $H_0$ with a bound less
than one, then $V$ is $H_\omega$-bounded with a bound less than
one too.
Therefore, $H_\omega+V$ is self-adjoint on $\cD(H_\omega +V)=\cD(H_\omega)$.
\end{lemma}
\begin{proof}
For the first part see \cite{Tip}. For $\Psi \in \cS(\R^\nu)$, a core for
$H_\omega$,
\begin{equation*}
\|V\;\Psi\| = \| V\; G^1\;G^{-1}\;\Psi\|\leq a\|H_0\;G^{-1}\;\Psi\| +
b\|G^{-1}\;\Psi\|
\leq a(1+\varepsilon)\|H_\omega \;\Psi \| + (a\,b(\varepsilon)+b)\|\Psi\|.
\end{equation*}
\end{proof}

The unitary propagator $\exp\{-i(t-s)(H_\omega
+ V )\}$ is \emph{formally} related to the propagator $U$ for the
time-dependent Schr\"{o}dinger equation \eqref{Ham} by
\begin{equation}\label{equality}
U(t,s)\ :=\ R(t)\; \exp\{-i(t-s)(H_\omega + V )\}\; R(s)^\ast.
\end{equation}
If $V$ is sufficiently smooth with respect to the angle $\phi$ then one
can verify that $U(t,s)$ maps a core into $\cD(H_0)$ and thus solves the
Schr\"{o}dinger equation with time-dependent Hamiltonian \eqref{Ham}.
However, even
without the additional smoothness when it is not so clear in which sense the
Schr\"{o}dinger equation is satisfied due to domain problems one should use the
propagator \eqref{equality}. It is justified by the discussion of rotating
frames and \eqref{rotframe} above. Next we will prove existence and
completeness of the wave operators \eqref{4}.

\section{Wave and Scattering Operators}
\setcounter{equation}{0}
In the inertial frame we have chosen
time $s=0$ as reference time for the wave operators $\Omega^\pm =
\Omega^\pm(H(t),\,H_0)$ in \eqref{4}. For another reference time $s$ one
has
\begin{equation}\label{e4.1}
\Omega^\pm_{[s]} =\slim_{t\to\pm\infty} U(t+s,\,s)^*\;
e^{-itH_0} = U(s,0)\;\Omega^\pm\;e^{isH_0}.
\end{equation}
There is no evident
intertwining relation between Hamiltonians because of the explicit
time dependence but due to periodicity we have it for monodromy operators:
\begin{equation*}
U(2\pi\omega^{-1}+s,\,s)\;\Omega^\pm_{[s]} = \Omega^\pm_{[s]}\;
\exp\{-i2\pi\omega^{-1}\,H_0\}.
\end{equation*}
See, however, \eqref{e4.5} below.
The corresponding scattering operators satisfy
\begin{equation*}
S_{[s]} = (\Omega^\pm_{[s]})^*\,\Omega^\pm_{[s]} = e^{-isH_0}\;
S_{[0]}\; e^{-isH_0}.
\end{equation*}
In general, they will depend on $s$ because the scattering operator needs
not commute with $H_0$.

We can combine the unitary families in \eqref{e4.1} differently to obtain
the evolutions in the rotating frame.
\begin{align*}
U(t+s,\,s)^* \; e^{-itH_0} &=
R(s)\: e^{it(H_\omega + V)} \: R(t+s)^*\; e^{-itH_0} = R(s)\:
e^{it(H_\omega + V)}
\;e^{-itH_\omega}\;R(s)^*\\
&=e^{it(H_\omega + V_s)}\;e^{-itH_\omega}
\end{align*}
where we have used $R(s)\,V\,R(s)^*=V_s$ in the last equality.
Different wave operators are thus related by
\begin{equation*}
\Omega^\pm_{[s]} = R(s)\;\Omega^\pm (H_\omega+V,\,H_\omega)\;R(s)^*
= \Omega^\pm (H_\omega+V_s,\,H_\omega).
\end{equation*}
Instead of comparing the standard free time evolution with a perturbed one
which has a time-dependent rotating potential one can study equivalently
the more complicated unperturbed evolution in the rotating frame and
its perturbation by a time-independent potential. If these wave operators
exist we immediately get the intertwining relation
\begin{equation}\label{e4.5}
e^{-i\tau (H_\omega+V_s)}\; \Omega^\pm (H_\omega+V_s,\,H_\omega) =
 \Omega^\pm (H_\omega+V_s,\,H_\omega)\;
e^{-i\tau H_\omega},\quad \tau\in\R\,.
\end{equation}
Now we can apply results of standard scattering theory. We consider first the
time-independent formulation in the rotating frame and we treat the physical case of
dimension $\nu=3$ as an example. The assumption on the decay of the potential is
fulfilled if, e.g., $|V(\x)| \sim |\x|^{-\beta}\;$ as $\,|\x| \to \infty$, $\beta >
7$.

\begin{theorem}\label{wav}
Let the potential $V$ satisfy $(1+|\x|^2)^2\,V\in
L^1(\R^3)\cap L^2(\R^3)$. Then the wave
operators $\Omega^\pm(H_\omega+V,\,H_\omega)$ exist and are complete,
$\Ran\, \Omega^\pm(H_\omega+V,\,H_\omega) = \cH_{\mathrm{ac}}(H_\omega+V)$.
\end{theorem}

\begin{proof}
\begin{eqnarray*}
|V|^{1/2}\;(H_\omega+i)^{-1} &=& |V\;G^2\,|^{1/2}\; G^{-1}\;
(H_\omega+i)^{-1}\\ &=&
|V\;G^2\,|^{1/2}\;(H_0+1)^{-1}\cdot (H_0+1)\; G^{-1}\;(H_\omega+i)^{-1}.
\end{eqnarray*}
By Lemma \ref{lemma:Tip} and since $\cS$ is a core for $H_\omega$ we have
that $(H_0+1)\; G^{-1}\;(H_\omega+i)^{-1}$ defines a bounded operator.
Further we estimate
\begin{equation*}
\|\ |V\;G^2\,|^{1/2}\;(H_0+1)^{-1}\|_{HS}^2 \leq
\mathrm{const}\, \|V\;G^2\|_{L^1}
\end{equation*}
which is finite by assumption.
Thus, $|V|^{1/2}\:(H_\omega+i)^{-1}$ is Hilbert-Schmidt.

We prove now that $|V|^{1/2}\:(H_\omega + V +i)^{-1}$ is also
Hilbert-Schmidt. To this end we use the resolvent equation and write
\begin{equation*}
|V|^{1/2}\;(H_\omega + V +i)^{-1} = |V|^{1/2}\;(H_\omega+i)^{-1}-
|V|^{1/2}\;(H_\omega+i)^{-1} V\; (H_\omega + V +i)^{-1}.
\end{equation*}
Since $V$ is $H_\omega$-bounded with bound less than one,
the operator $V\; (H_\omega + V +i)^{-1}$ is
bounded, and thus, $|V|^{1/2}\;(H_\omega + V +i)^{-1}$ is Hilbert-Schmidt.

Since $|V|^{1/2}\;(H_\omega+i)^{-1}$ and $|V|^{1/2}\;(H_\omega + V +i)^{-1}$
are both Hilbert-Schmidt we apply the resolvent equation to obtain that
$(H_\omega + V +i)^{-1}-(H_\omega+i)^{-1}$ is trace class. By the
Kuroda-Birman theorem \cite{RS3,Yafaev} existence and completeness of the
wave operators follows. This completes the proof of Theorem \ref{wav}.
\end{proof}

Obviously the same applies to wave operators for any other reference time $s$, i.e.,
if one replaces $V$ by $V_s$. We state now the result in the setting of rotating
potentials. The absolutely continuous spectral subspaces then correspond to the
monodromy operators for one period $2\pi/\omega$.
\begin{corollary}\label{cor4.3}
Let the potential satisfy
$(1+|\x|^2)^{\alpha}\;V\in L^1(\R^3)\cap L^2(\R^3)$ for some $\alpha\geq 2$.
For any reference time $s$ the wave operators $\Omega^\pm_{[s]}$ given by
\eqref{e4.1} exist and are complete in the sense that
\begin{equation*}
\Ran\ \Omega^\pm_{[s]} =  \cH_{\mathrm{ac}}(H_\omega+V_s)=R(s)\;
\cH_{\mathrm{ac}}(H_\omega+V) =
\cH_{\mathrm{ac}}(U(s+2\pi/\omega,\,s)).
\end{equation*}
The scattering operator $S_{[s]}$ is unitary and $H_\omega$ is conserved under
scattering:
\begin{equation*}
e^{-i\tau H_\omega}\;S_{[s]} = S_{[s]}\; e^{-i\tau H_\omega},\quad \tau\in\R\,.
\end{equation*}
If, in addition, the distributional azimuthal derivative of the potential is bounded
and satisfies $(1+|\x|^2)^\beta\; \partial_\phi V \in L^1(\R^3)\cap L^2(\R^3)$ for
some $\beta > 1/2$ then the kinetic energy is uniformly bounded.
\end{corollary}
\textbf{Remark.}
The boundednes of $\partial_\phi V$ has been assumed in Section~2 for
simplicity of presentation. This condition can be relaxed for rotating
potentials e.g.\ to $\|\partial_\phi V\; (H_\omega +i)^{-1}\|<\infty$
or $\|(1+|\x|^2)\;\partial_\phi V\; (H_0 +1)^{-1}\|<\infty$, cf.\
the proof of Proposition~\ref{erste:Proposition} and Lemma~\ref{lemma:Tip}.
\begin{proof}
The first condition on the potential is the assumption of Theorem~\ref{wav}.
It ensures that condition
\eqref{19} of Lemma~\ref{lem2.2} is satisfied: The free resolvent is a bounded
map $L^2(\R^3) \to L^\infty(\R^3)$. Therefore $(1+|\x|^2)^2\;V\; (H_0 +
1)^{-1}$ is a bounded operator on $L^2$.
This implies boundedness of $\,\|V\;(H_0 +1)^{-1}\;(1+|\x|^2)^2\|$
because the resolvent acts in
configuration space as a convolution with a continuous rapidly decaying
function. In particular, \eqref{19} follows.
Similarly, the assumption in the last statement implies that \eqref{22} is
satisfied as well.
\end{proof}

\section{Scattering off a Rotating Blade}
\setcounter{equation}{0}

In this section we give a rough approximate description of energy transfer when a
microscopic quantum particle hits a rotating macroscopic reflecting blade. During the
scattering process the wave packet is assumed to be small compared to the size of the
blade and the separation of the collision point from the axis of rotation. In
addition, the speed of the collision point on the blade is small compared to the
speed of the quantum particle (small $\omega$) and the transmission through the blade
by tunnelling is negligible.

As in Section~3 we construct suitable states starting from a rotational
symmetric $\Psi_0$ which has a smooth compactly supported momentum space
wave function. It has zero angular momentum
$J\,\Psi_0 =0$. This time
\begin{equation*}
\widehat{\psi}_{b,\vitsmall}(\p)\ :=\ e^{-ibp_2}\;\widehat{\psi}_0 (\p + m\vit\e_1)
\end{equation*}
describes a state which moves with velocity $-\vit$ in the $\e_1$-direction and
is localized in configuration space near $x_2=b$, $x_j \approx 0$ for $j\neq
2$. To ensure good propagation properties we assume that
$\supp\,\widehat{\psi}_0 \subset B_{m\vitsmall/3}(\mathbf{0})$. Such a state
has impact parameter $b$ and it is localized in angular momentum space near $-b
m \vit$. In our units of measurement where Planck's constant $\hbar=1$ we have
$|-b m \vit|\gg 1$ for a macroscopic impact parameter and e.g.\ thermal
velocities. Therefore the quantization of angular momentum is not relevant
here.

The blade is represented by a strong potential with support near the hyperplane through the origin
perpendicular to the $\e_1$-direction, e.g., in two dimensions near the line $x_1=0$, $|x_2|\leq
B$.

In the past the state has been essentially localized under the free time
evolution far away from the support of the potential and it was ``incoming''
from the right. Superimposing the rotation does not change the good separation
from the potential if the parameters are suitably chosen, namely $B\omega$
small enough compared to $\vit$. We use this to show that in good approximation
the wave operator can be calculated using a small finite negative time
$-\sigma$ when the scattering sets in: $\Omega^- (H_\omega +V,\,H_\omega)
\approx \exp\{i(-\sigma) (H_\omega +V)\}\;
\exp\{-i(-\sigma) \, H_\omega\}$.

Let $\chi_{G(t)}$ denote the characteristic function in configuration space of
a region $G(t)\subset\R^\nu\,$ and $\widetilde{\chi}_{G(t)}$ its convolution
with a smooth function with integral one and support in a ball of radius one.
Then $\nabla \widetilde{\chi}_{G(t)}$ and $\Delta \widetilde{\chi}_{G(t)}$ are
uniformly bounded and have support in the union of balls $B_1(\partial G(t))$.
The same holds for $1-\widetilde{\chi}_{G(t)}$. The function
$\widetilde{\chi}_{G(t)}$ is supported in $B_1(G(t))$ while the support of
$1-\widetilde{\chi}_{G(t)}$ is contained in $B_1(\R^\nu\setminus G(t))$. We
choose the family $G(t)$ for negative times such that the main part of the
state $e^{-itH_\omega}\;\Psi_{b,\vitsmall}$ is localized inside $G(t)$ and
$\lim_{t\to
-\infty}[\,1 - \widetilde{\chi}_{G(t)}\,]\; e^{-itH_\omega}\;\Psi_{b,\vitsmall}=0$.
Then
\begin{align}\label{e5.2}
&\Omega^- (H_\omega +V,\,H_\omega)\;\Psi_{b,\vitsmall}
-e^{i(-\sigma) (H_\omega +V)}\;\widetilde{\chi}_{G(-\sigma)}\;
e^{-i(-\sigma) \, H_\omega}\; \Psi_{b,\vitsmall} \notag \\[1ex] &=
\lim_{T\to -\infty}\left\{e^{iT (H_\omega +V)} \;
\widetilde{\chi}_{G(T)}\; e^{-iTH_\omega}
- e^{i(-\sigma) (H_\omega +V)}\;\widetilde{\chi}_{G(-\sigma)}\;
e^{-i(-\sigma) \, H_\omega} \right\} \Psi_{b,\vitsmall}
\end{align}
If for $t\leq -\sigma$ the condition $\supp\, V \cap
\supp\,\widetilde{\chi}_{G(t)}=\emptyset$
is satisfied then the r.h.s.\ can be estimated by
\begin{align}\label{e5.3}
&\int_{-\infty}^{-\sigma} dt\;\left\|\frac{d}{dt} e^{it (H_\omega +V)}
\;\widetilde{\chi}_{G(t)}\; e^{-itH_\omega}\;\Psi_{b,\vitsmall} \right\| \notag
\\[1ex]
&\leq \int_{-\infty}^{-\sigma} dt\;\Bigl\{\left\|\, [\,H_\omega
,\,\widetilde{\chi}_{G(t)}\,]\;e^{-itH_\omega}\;\Psi_{b,\vitsmall}\right\| +
\left\|\,(\partial_t \widetilde{\chi}_{G(t)})\,e^{-itH_\omega}\;\Psi_{b,\vitsmall}
\right\|\Bigr\} \notag \\[1ex]
&\leq \mathrm{const} \int_{-\infty}^{-\sigma} dt\; \Bigl\{\left\| F\{\x\in
B_1(\partial G(t)\,)\}\;e^{-itH_\omega}\;\Psi_{b,\vitsmall}\right\| + \left\|
F\{\x\in B_1(\partial
G(t)\,)\}\;e^{-itH_\omega}\;\p\:\Psi_{b,\vitsmall}\right\|\Bigr\}
\end{align}
where the constant takes care of the suprema of first and second derivatives
of $\widetilde{\chi}_{G(t)}$ which are independent of $t$ and $F\{\x\in M\}$
is the multiplication operator in configuration space with the characteristic
function of $M$. Since
\begin{equation*}
\|F\{\x\in B_1(\partial G(t)\,)\}\;e^{-itH_\omega}\;\Psi_{b,\vitsmall}\| =
\|F\{\x\in \cR(t)\,B_1(\partial G(t)\,)\}\;e^{-itH_0}\;\Psi_{b,\vitsmall}\|
\end{equation*}
we can apply the propagation estimate \eqref{14} for the free time evolution.

We choose
\begin{equation*}
G(t) = \cR(t)^{-1}\;B_{\rho+1+|t|\vitsmall/2}(b\e_2 - t\vit\e_1).
\end{equation*}
Then $\|F(\x\in B_1(\partial G(t)\,)\;e^{-itH_\omega}\;\Psi_{b,\vitsmall}\|\leq
\mathrm{const}\;(1+\rho+|t|)^{-2}$. The same estimate applies to the term with
$\p\;\Psi_{b,\vitsmall}$. The integral \eqref{e5.3} is as small as desired by
choosing $\rho$ large enough. The support of the potential is separated by 1
from the support of $\widetilde{\chi}_{G(t)}$ for all small enough $\omega$ and
times $t<-\sigma:=-2(\rho+3)/\vit$. The approximation of the incoming wave
operator as given on the l.h.s.\ of
\eqref{e5.2} is as good as needed. Moreover,
\begin{equation*}
\sup_{-\sigma\, <\, t\, <\, 0} \left\|\left(e^{it (H_\omega +V)} \;
\widetilde{\chi}_{G(t)}\; e^{-itH_\omega} -
e^{it (H_\omega +V)} \; e^{-itH_\omega}\right) \;\Psi_{b,\vitsmall} \right\| =
o(\rho)
\end{equation*}
is small as well. An analogous estimate can be given for the outgoing wave
operator on suitably selected states and we obtain for the scattering
operator $S=S_{[0]}$:
\begin{align}
S\; \Psi_{b,\vitsmall} &\approx  e^{i\sigma \, H_\omega}\;
\widetilde{\chi}_{G(-\sigma)}\;
e^{-i2\sigma (H_\omega+V)}\; \widetilde{\chi}_{G(-\sigma)}\; e^{i\sigma \,
H_\omega}\; \Psi_{b,\vitsmall}\nonumber
\\[1ex]\label{e5.8}
&\approx e^{i\sigma \, H_\omega}\; e^{-i2\sigma (H_\omega+V)}\; e^{i\sigma \,
H_\omega}\; \Psi_{b,\vitsmall}
\end{align}
The approximation \eqref{e5.8} shows that the potential may be changed
arbitrarily far away from $G(t)$. In particular, it may be replaced by
a simpler potential barrier in the $x_1$-direction which is independent of the
other coordinates. Since the time interval $[-\sigma,\,\sigma]$ is bounded we
ignore $\omega t$ for small $\omega$ and we may replace the high potential
barrier by a Dirichlet boundary condition at $x_1=0$. For this Hamiltonian --
denoted by $H_D$ -- the eigenfunctions on $\R^\nu_+ = \{\x\in\R^\nu\mid
x_1\geq 0\}$ are
\begin{equation*}
e^{i\p\x} - e^{i(\p\x - 2p_1x_1)},\quad x_1\geq 0.
\end{equation*}
In this approximation $S$ acts as a reflection at the hyperplane $x_1=0$ in the
rotating frame.

We know from Corollary~\ref{cor4.3} that $H_\omega = H_0 - \omega J$ is conserved under scattering
but the angular momentum of $\Psi_{b,\vitsmall}$ changes sign under reflection $-bm\vit\to
bm\vit$. Consequently, the kinetic energy changes,
\begin{equation*}
S\; H_0\; \Psi_{b,\vitsmall} = S\; (H_\omega + \omega J)\;\Psi_{b,\vitsmall} \approx (H_\omega -
\omega J)\; S \;\Psi_{b,\vitsmall} = (H_0 - 2\omega J)\; S \;\Psi_{b,\vitsmall} \approx (H_0 -
2\omega bm\vit)\; S \;\Psi_{b,\vitsmall}.
\end{equation*}
The energy increases for $\omega < 0$ when the relevant part of the blade
moves towards the particle. The behavior for quantum particles is the same
as for classical elastic balls.

For simplicity we have assumed an orthogonal collision. The energy
transfer is the same for other angles as long as the impact parameter $b$
remains unchanged. It determines the classical angular momentum. We will give
a better approximation with detailed error bounds in a forthcoming paper.


\end{document}